\def\chisq{$\chi^{2}_{\nu}$}\def\minone{$^{-1}$}\def\mintwo{$^{-2}$}
\def\sqiggt{\hbox{\rlap{\lower.55ex \hbox {$\sim$}}
\kern-.3em \raise.4ex \hbox{$>$}\,}} \def\kev{\,ke\kern-.1em V}
\def\sqig{$\sim\,$} \def\etal{et\,al.} \def\msun{M$_{\scriptstyle\odot}$} 
\def\up#1{$^{\mbox{{\scriptsize #1}}}$} \def\cps{counts s$^{-1}$} 
\def\lo#1{$_{\mbox{{\scriptsize #1}}}$} \def\pten#1{$\times10^{#1}$}
\def\deg{$^{\circ}$}  \def\mdot{$\dot{M}$}
\documentstyle{mn}
\title[An outburst of XY Ari]
{An outburst of the magnetic cataclysmic variable XY~Arietis observed with 
{\sl RXTE}}
\author[C.~Hellier \etal]{Coel Hellier,$^{1}$ Koji Mukai$^{2}$ and
A.~P.~Beardmore$^{1}$\\
$^{1}$Department of Physics, Keele University, Keele, Staffordshire, ST5 5BG\\
$^{2}$Laboratory for High Energy Astrophysics, Code 660.1, NASA/GSFC,
Greenbelt, MD 20771, USA}

\date{ }
\begin{document}
\maketitle
\begin{abstract}
We report the first observed outburst of the magnetic cataclysmic variable 
XY~Ari. X-ray observations show a flux increase by an order of magnitude the
day after the first signs of outburst. During the 5-d duration the X-ray
spin pulse is greatly enhanced and the X-ray spectrum far more absorbed. We
suggest that the inner disc pushes inwards during outburst, blocking the
view to the lower accreting pole, breaking the symmetry present in
quiescence, and so producing a large pulsation. The observations are
consistent with a disc instability as the cause of the outburst, although we
can't rule out alternatives. We draw parallels between our data and the UV
delay and dwarf nova oscillations seen in non-magnetic dwarf novae.
\end{abstract}
 
\section{Introduction}
The dwarf nova (DN) eruption is one of the most distinctive features of the
cataclysmic variables (CVs). Such stars brighten by typically 3--5 magnitudes
for durations of 4--8 d every 20--60 d. There are also longer, brighter
`superoutbursts' and other related phenomena (for a comprehensive review see
Warner 1995a). The near unanimous opinion is that the outbursts are caused by
instabilities in the accretion disc surrounding the white dwarf in these close
binaries. Because of the hydrogen ionization instability a disc with the right
conditions can choose either of two states (one hot and viscous, the other
cooler and more fluid) for a given surface density. The disc cycles between the
states, dumping material onto the white dwarf at a higher rate when it is
hot and viscous (reviewed by Osaki 1996). The alternative
proposal, that instabilites in the secondary star increase the mass transfered
into the disc, is decreasingly favoured for typical DN eruptions, but
might still have a role to play in explaining the full phenomenology of CVs.

While outbursts in non-magnetic DNe are well studied, far fewer have
been seen in magnetic systems. In AM~Her stars, which don't have discs,
the lack of outbursts is expected. The intermediate polars (IPs), which 
have weaker fields and are not phase-locked, but do possess at least
partial discs, show several variations on the DN theme. These are 
reviewed below to set the scene for this paper.

XY~Ari is an IP lying behind a molecular cloud (Patterson
\&\ Halpern 1990) and so cannot be seen in the optical. X-ray and infra-red
studies (Kamata, Tawara \&\ Koyama 1991; Zuckerman \etal\ 1993; Allan,
Hellier \&\ Ramseyer 1996) show it have a 6.06-h orbital period and a 206-s 
spin period, and reveal deep eclipses implying an inclination of $>$80\deg.
To use the X-ray eclipse to study the accretion regions we observed
20 eclipses of XY~Ari with {\it RXTE\/} (Bradt, Rothschild \&\ Swank 1993).
The results are reported in Hellier (1997a). However, XY~Ari went into 
outburst during the observations, the first ever seen
of this star. It is also only the second magnetic CV to have
been observed in outburst in X-rays, after GK~Per (Watson, King \&\ Osborne
1985). This paper reports the outburst and discusses its
implications for IPs and other CVs.

\subsection{Outbursts in intermediate polars}
Table~1 lists the six well-established IPs which have 
shown outbursts, together with the relevant literature (see Hellier 1993a
and Warner 1997 for previous compilations). GK~Per has a long orbit and an
evolved secondary, and its month-long outbursts can be explained as disc
instabilities when the parameters are adjusted for GK~Per's peculiarities (Kim,
Wheeler \&\ Mineshige 1992). YY~Dra, and now XY~Ari, have shown outbursts lasting
5 days, which is typical of DNe. The recurrence time for YY~Dra's
outbursts (several years) is an order of magnitude greater than typical. 
However, the inner disc is missing in an IP, and, since
the disc thus needs to disipate less angular momentum, its outer radius can
be less than in non-magnetic systems. The truncated disc takes longer to 
accumulate sufficient material for an outburst (Angelini \&\ Verbunt 1989; 
Warner 1997). The latter paper predicted that XY~Ari should have outbursts
similar to those of YY~Dra, which this paper shows to be correct.

\begin{table}
\caption{Intermediate Polars showing outbursts}
\begin{center}\begin{tabular}{lccccr}\\
Star & Length  &  $\Delta$mag & Interval & P\lo{orb} & P\lo{spin} \\ 
         &   (day)  &    &            &  (hr)  & (secs) \\ [3mm]
V1223~Sgr &   0.5      &    $>$1     &     ?      &  3.37 & 745 \\ [2mm]
TV~Col   &    0.5      &    2        & \sqig1 month? & 5.49 & 1910 \\ [2mm]
EX~Hya   &    2      &    4        & \sqig yrs, erratic  &  1.64 & 4022 \\ [2mm]
XY~Ari   &    5      & \sqig3      &   \sqiggt 50 d  &  6.06 & 206  \\ [2mm]
YY~Dra   &    5      &      5      &  \sqig 1000 d  & 3.96 & 529 \\ [2mm]
GK~Per   &    50     &    3        & 880--1240 d & 47.9 & 351 \\ [3mm]
\end{tabular} \end{center}
\parbox{85mm}{\footnotesize Refs: V1223~Sgr: van Amerongen \&\ van Paradijs 
1989. TV~Col: Szkody \&\ Mateo 1984; Schwarz \etal\ 1988; Hellier \&\ Buckley
1993. EX~Hya: Bateson \etal\ 1970; Hellier \etal\ 1989; Reinsch \&\ 
Beuermann 1990; Buckley \&\ Schwarzenberg-Czerny 1993. XY~Ari: This paper.
YY~Dra: Patterson \etal\ 1992; Mattei 1996b. GK~Per: Watson \etal\ 1985; 
Bianchini \etal\ 1986; Kim \etal\ 1992; Mattei 1996a; Morales-Rueda 
\etal\ 1996.} \end{table}

Note that the three stars mentioned so far have spin periods 
less than 5 per cent of the orbital periods, atypically low amongst IPs. 
This means that their magnetospheres are small in comparison with the discs,
allowing the discs to behave similarly to those in non-magnetic systems.

The remaining three stars showing outbursts, EX~Hya, TV~Col and V1223~Sgr, are
among the majority of IPs with spin periods greater than 5 per cent of the
orbital period, where much more of the inner disc is magnetically disrupted.
These stars present more of a problem for the disc instability model. While
their low amplitudes, short durations, and long recurrence times can to some
extent be explained by disc trunction (Angelini \&\ Verbunt 1989), the small
fraction of the transfered material involved in the outbursts is atypical, and
suggests that disc instabilities are suppressed, with the observed outbursts
having a different cause (Warner 1997).

Hellier \etal\ (1989) and Hellier \&\ Buckley (1993) proposed that the
outbursts in these three stars result from increased mass transfer from the
secondary. The evidence for this (or at least against the disc-instability
model) can be summarised as: (1) In all three stars the line emission increases
during outburst, in contrast to normal DN behavior. (2) In both EX~Hya 
and TV~Col the line emission from the bright-spot where the stream hits the 
disc increases relative to the rest of the system during outburst, suggesting 
increased mass transfer from the secondary. (3) EX~Hya shows the stream
oveflowing the disc in outburst, but not in quiescence, a possible consequence 
of an enhanced stream. (4) The optical eclipse in EX~Hya is centered later in
outburst than quiescence (Reinsch \&\ Beuermann 1990) also suggesting an 
enhanced stream. (5) The eccentric disc in TV~Col shows a phase
glitch in outburst, as  expected if it encounters a large amount of new
material. (6) V1223~Sgr shows VY~Scl low states (Garnavich \&\ Szkody 1988)
in addition to outbursts. Since VY~Scl stars in their high state are on
the hot side of the disc instability, according to the usual picture,
the outbursts seen on top of the high states can't be disc instabilities.
(7) EX~Hya has shown two outbursts separated by 8 days, and also periods
of $>$3 y without outbursts, but no difference in the quiescent magnitude.
This is hard to account for in the disc instability model.

In addition, there is substantial evidence that similar short, low-amplitude 
outbursts occur in other types of CV, often hidden amongst the disc
instabilities. Bateson (1991) lists many such flares in non-magnetic DNe; a
particularly well studied example is presented by Echevarria \etal\ (1996),
who interpret it as a mass transfer event. Further, O'Donoghue (1990) saw
one anomalous eclipse profile of Z~Cha early in a superoutburst, which
implied a bright spot 20 times more luminous than in quiesence. There  has
also been a short-lived flare from the AM~Her star QS~Tel observed in the
EUV (Warren \etal\ 1993). Since AM~Her stars don't have discs this can only
have been a mass-transfer event or a secondary star flare.
AM~Her stars also have low states (e.g.~Cropper 1990), which must be caused 
by mass-transfer variations, although the timescales (typically
months) are longer than the flares and outbursts discussed here.

\begin{figure*}\vspace{8.3cm}   % Fig 1
\caption{The 20 {\it RXTE\/} lightcurves of XY~Ari. Day zero (here and in
Fig.~2) is HJD 245\,0267.8525 or 1996 July 3, 8:32 {\sc ut}.}
\includegraphics{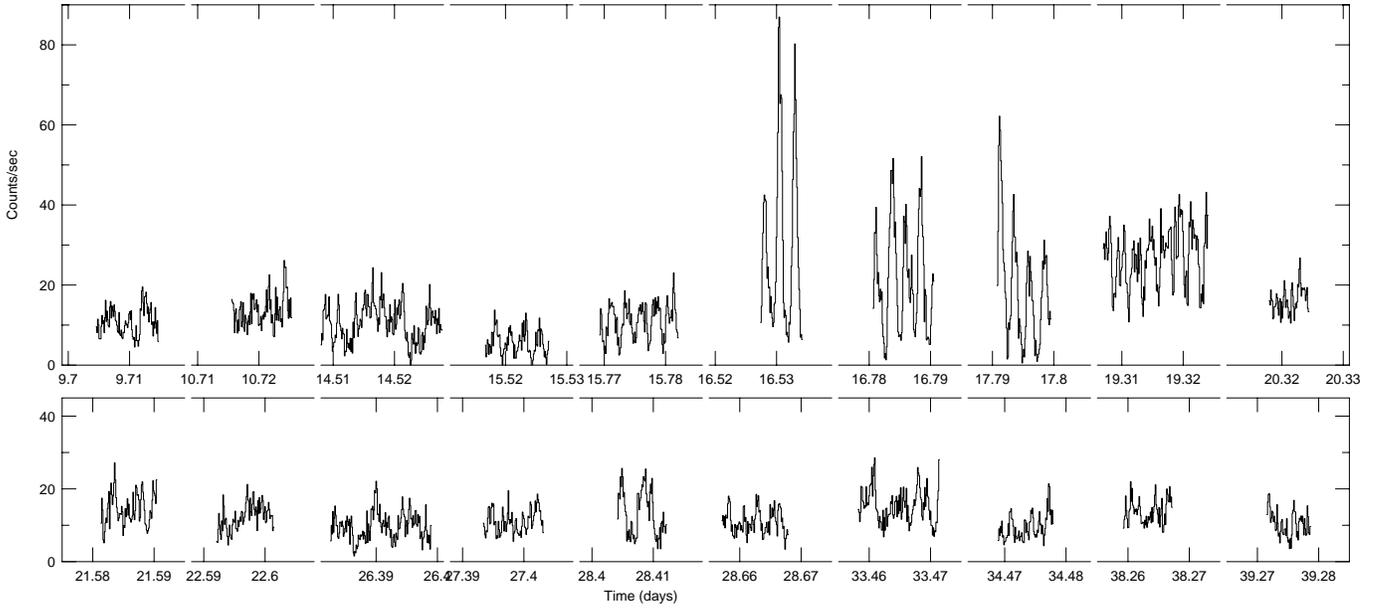}
\end{figure*}

\section{Observations}
The 20 eclipse egresses were observed by {\it RXTE\/} over the period 1996 July
13 to 1996 August 11, at a rate of roughly one per day. Each observation lasted
\sqig2000 s, with the first 1000 s in eclipse and the later half covering
\sqig5 spin cycles. For this paper we discard the eclipse and egress
information (for which see Hellier 1997a) and concentrate on the outburst. 

We extracted lightcurves in the energy range 2--15 \kev, using data from the
top Xenon layer of the PCA only [this gives a higher signal-to-noise (S/N)
for dim sources]. We then subtracted the background estimated using {\sc
pcabackest} v1.4f. This marginally underestimated the count rate in eclipse,
so, as a final step, we fitted a straight line to the data in eclipse, 
extrapolated it through the observation, and subtracted it to give zero
counts in eclipse. 
 
The 20 lightcurves are presented in Fig.~1. Note that in reality the
gaps between the observations (\sqig 1 d) are large compared to the 
length of each observation (\sqig 15 min), but they are highly contracted
to display the whole dataset in one plot.

\begin{figure*}\vspace{12cm}   % Fig 2
\caption{The top panel shows the average count rate in each of the 20
observations. The next panel show the amplitude of the pulsation in each
observation, obtained by fitting the folded data with a sinusoid and 
recording the full amplitude of the sinusoid divided by the peak value.
The third and bottom panels show the results of fitting the spectrum
from each observation with a simple model of a 30 \kev\ bremsstrahlung plus
absorption. The absorption column (third panel) is in units of $10^{22}$
cm\mintwo, while the bottom panel shows the relative normalisation of the
bremsstrahlung emission. Formal photon-noise error bars are comparable to the
size of the plotted symbols. However, in all cases uncertainty due to
flickering or model ambiguity is greater than errors derived from photon 
statistics.}
\includegraphics{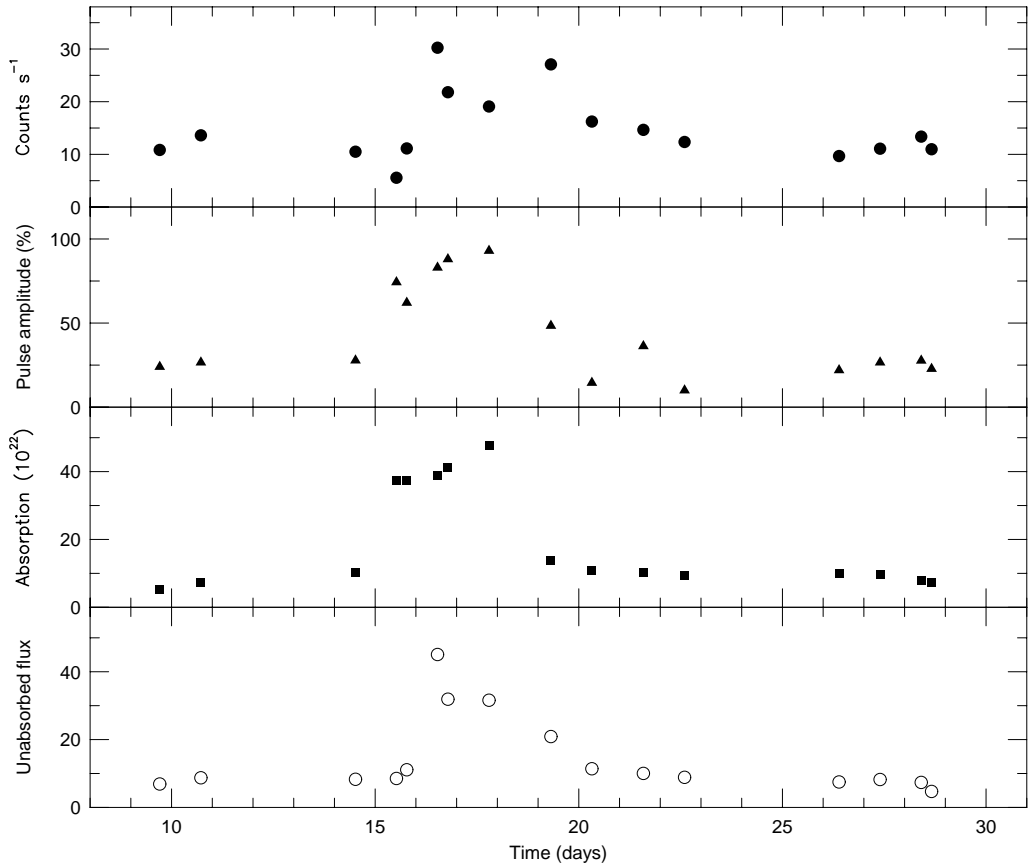}
\end{figure*}

\section{The spin pulse}
The most striking aspect of Fig.~1 is that  while the mean count rate increases
during the outburst, the  amplitude of the spin pulse increases from a
quiescent level of 20--30 per cent (so low that it is hard to see amongst the
flickering) to 80--90 per cent. Furthermore the increase in pulse amplitude
occurs a day before the average count rate rises. This is illustrated in
Fig.~2, which shows the mean count rate and pulse amplitude, plotted, this 
time, on a continuous time axis. 

Fig.~3 compares the pulse profiles at all stages through the outburst. The
quiescent pulse (the average of all 14 observations before and after the
outburst) is shown at the bottom and top of the plot. It has a low amplitude
and is double-peaked (as it was in the {\it Ginga\/} observations reported by
Kamata \&\ Koyama 1993). The 6 outburst observations (observations 4 to 9 in
Fig.~2) follow in order from the bottom upwards. In outburst the profile 
becomes single peaked, and at the peak of the outburst it is nearly sinusoidal.
However, both on the rise and fall (observations 5 and 9) the profile is flat
topped. 

\begin{figure}\vspace{14.5cm}   % Fig 3
\caption{The evolution of the spin pulse through the outburst. The folded
data from all 14 quiescent observations are shown both at the top and the 
bottom. The pulse profiles from the outburst observations are shown
with time increasing upwards. Each pulse has been normalized to have a 
mean value of 1 and an offset of 1 is added to each successive profile
for clarity. Thus the zero point is at zero for the first profile, and
increases by 1 each time. Again, photon-noise errors are small (see Fig.~4),
and the dominant source of uncertainty is flickering.}
\includegraphics{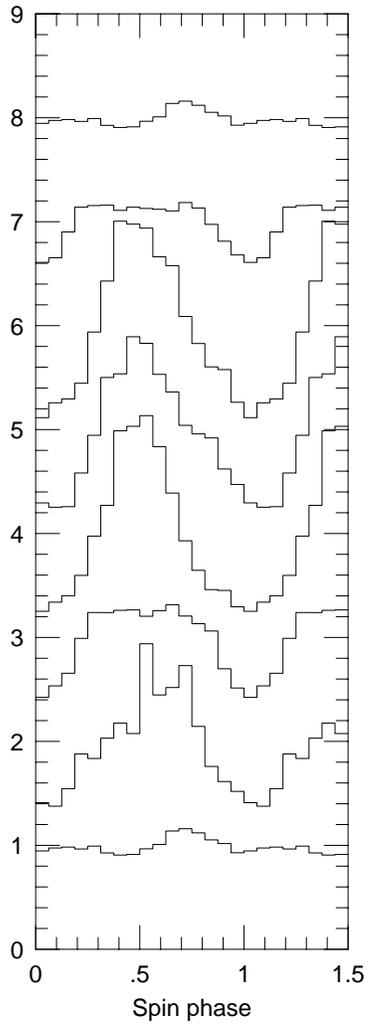}
\end{figure}

\begin{figure}\vspace{12.5cm}   % Fig 4
\caption{The upper two panels show the average spin pulse from all 6
outburst observations, together with the 6--15/2--5 \kev\ hardness ratio.
The lower panels are the same for the 14 quiescent observations. The 
spin phase is defined so that the upper magnetic pole is in the middle
of the white dwarf face at phase 0.5 (see text).}
\includegraphics{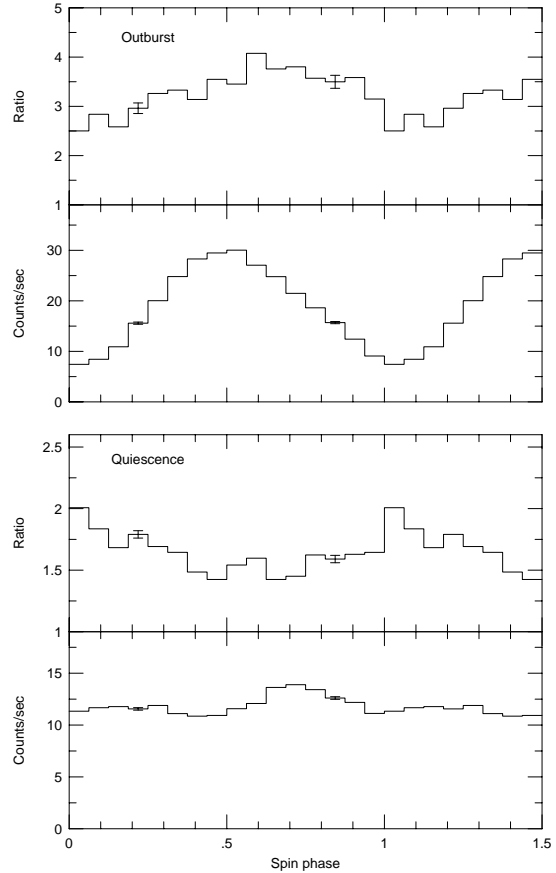}
\end{figure}

The mean quiescent and outburst pulse profiles are shown together with the
6--15/2--5 \kev\ hardness ratios in Fig.~4. The hardness ratio increases
markedly in the outburst, from \sqig1.7 to \sqig3. Further, during outburst,
the hardness ratio varies with spin phase, being greater at flux maximum.

\section{The spectrum}
The quiescent spectrum is adequately fit by an absorbed 30 \kev\ bremsstrahlung,
although the adequacy derives largely from the limited spectral resolution
of {\it RXTE\/} and the low number of counts from such short observations.
As a check we extracted from the archive the 40 ks {\it ASCA\/} observation 
of XY~Ari performed on 1995 August 7. The spectral resolution of
{\it ASCA\/} is much better but the low count rate of 0.158 \cps\
limits the usefulness of the data. The 0.5--10 \kev\ SIS and GIS
spectra are again well fitted by a bremsstrahlung with a (fixed) temperature
of 30 \kev\ and an absorption column of 6\pten{22}\ H atom cm\mintwo. 
With the addition of an iron line at 6.6 \kev, this model gives \chisq\, = 1.0.

The spectra during outburst show greatly increased absorption, explaining the
increased hardness ratio (Fig.~5). To investigate this we fitted the spectra
from all 20 observations with the same absorbed 30 \kev\ bremsstrahlung model,
allowing the column and the normalisation of the bremsstrahlung to optimise
for each observation individually. The resulting parameters are plotted in the
lower panels of Fig.~2. While this model fits the quiescent data adequately
(\chisq\, = 1.0), it was a poorer fit to the outburst spectra (\chisq\, = 1.8). 

Unfortunately, the low spectral resolution of {\it RXTE\/} means that fitting
more complex models to the outburst spectra becomes ambiguous. They can
be fit by any of (1) adding a further, dense, partial-covering absorber;
(2) adding lower temperature bremsstrahlung or black-body components; or
(3) adding a broad iron line together with a deep iron edge.

We prefer the first model because it is consistent with the increased
absorption required whatever the model; because it gives the lowest \chisq\
overall; and because there is little sign of the other features in the
quiescent {\it RXTE\/} or {\it ASCA\/} spectra. Further, the dense partial
absorber is consistent with our interpretation of the spin pulse (see
Section~5). This model gives \chisq\, = 0.95 in a simultaneous fit to the
six outburst spectra plus the averaged quiescent spectrum.  The
bremsstrahlung temperature was fixed at 30 \kev\ while the column of the
partial absorber was common to all spectra, and gave a best fit value of
$170\pm40$ \pten{22}\ cm\mintwo. The covering fraction of the partial
absorber, the column of the simple absorber, and the normalization of the
bremsstrahlung were fitted to each spectrum individually, and the resulting
values are shown in Fig.~6. For this figure, and when estimating the
change in accretion rate (Section 5.3), we decreased the
normalisation of the quiescent spectrum by half, to account for the fact
that we are probably seeing both poles in quiescence but only one in
outburst (see Section~5). Thus, even though there is ambiguity in the model
fits, the values in Fig.~6 are likely to be a truer representation of the
outburst than those in Fig.~2.

\begin{figure}\vspace{8cm}   % Fig 5
\includegraphics{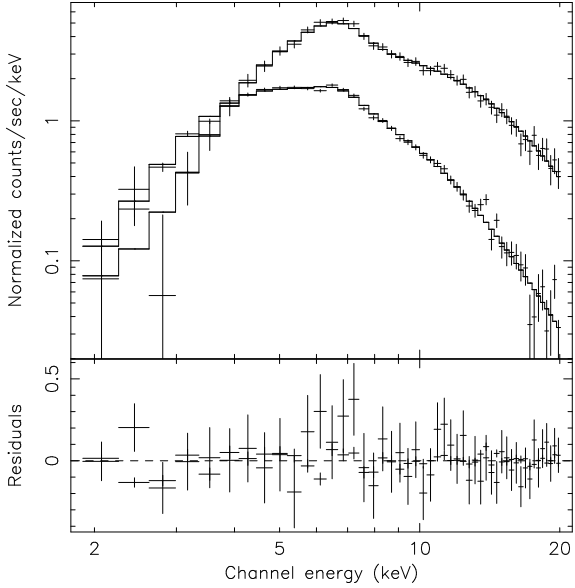}
\caption{The spectrum from the peak of the outburst (observation 6)
compared with the averaged quiescent spectrum. The quiescent spectrum
is fitted with an absorbed bremsstrahlung plus an iron line, while the
outburst model contains an additional partial absorber (see Section~4).
The observed fluxes are 3.1\pten{-11}\ (quiescence) and
11.5\pten{-11}\ (outburst) erg cm\mintwo\ s\minone\ over 2.0--20.0 \kev. 
The corresponding unabsorbed fluxes are 4.2\pten{-11}\ and 66.0\pten{-11}\ erg 
cm\mintwo\ s\minone\ respectively.}
\end{figure}

\begin{figure}\vspace{13cm}   % Fig 6
\includegraphics{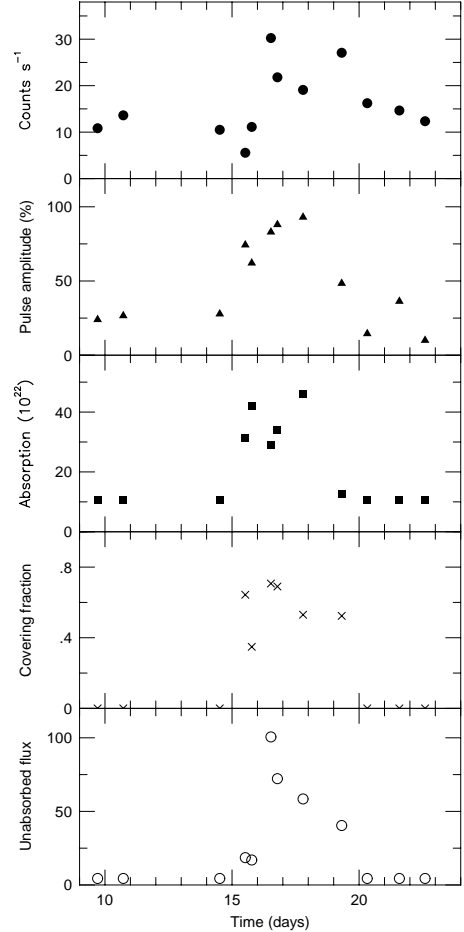}
\caption{A version of Fig.~2 but using a more complex absorption model.
The top two panels are the mean count rate and the pulse amplitude, as in
Fig.~2. The lower three panels result from fitting the data with a 30 \kev\
bremsstrahlung absorbed by both a simple absorber, whose column density is shown
in the middle panel, and a dense 1.7\pten{24}\ cm\mintwo\ partial absorber,
whose covering fraction is given in panel second from bottom. The bottom panel
shows the normalisation of the bremsstrahlung in this model. The quiescent
points all have the same value in the lower panels, since the model was fitted
to their average. Again, the uncertainty in the values results from the 
flickering in such short observations, and from the ambiguity in the fitted
models. Thus the photon noise error bars are in all cases small compared
to the scatter in the points.}
\end{figure}

\begin{figure}\vspace{8cm}   % Fig 7
\includegraphics{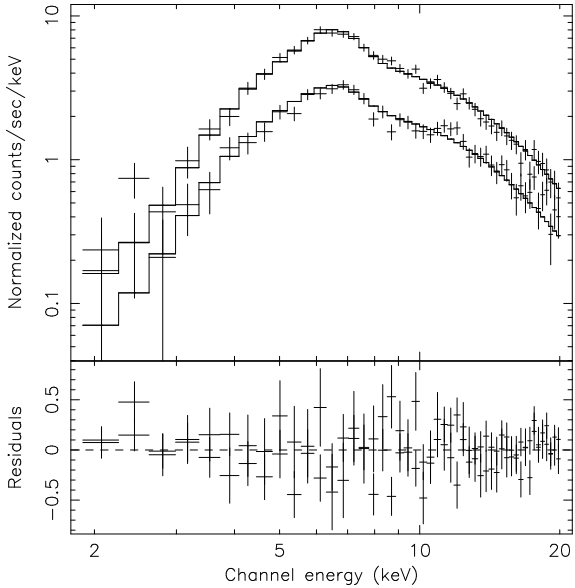}
\caption{The pulse maximum spectrum from the peak of the outburst (observation
6) compared with the pulse minimum spectrum from the same observation. The
fitted models are the same except for a change in normalization.}
\end{figure}

We have also investigated the spectral changes over the spin cycle in
the outburst observations (in quiescence the pulse amplitude and count rate are
too low for phase-resolved analysis). We found that the spin pulse is
almost purely a simple change in the intensity, with no accompanying change
in the spectrum. As an illustration Fig.~7 shows the data from the peak of
outburst (observation 6) binned into a spin-maximum spectrum 
(when the count rate was $>$60 per cent of the mean) and a spin-minimum
spectrum (count rate $<$40 per cent of the mean). The spectra were fitted 
with the same model of a 30 \kev\ bremsstrahlung and an iron line, absorbed by 
a 30\pten{22}\ cm\mintwo\ simple absorber and a 170\pten{22}\ cm\mintwo\
absorber  covering 0.60 of the source. Allowing only the normalization to
change between the spectra gives a \chisq\ of 0.98. The hardness ratio from all
outburst observations combined does show a slight change over the spin cycle,
getting harder at spin maximum (Fig.~4), but the phase resolved spectra from
individual observations are still adequately fit without including this effect.

\section{Interpretation}
To present our interpretation we need to review the results of the 
eclipse study in Hellier (1997a). This concluded that accretion occurred
onto two approximately equal poles at latitudes between 43\deg\ and 63\deg.
Each pole crosses the white dwarf face in half a spin cycle, at which point
its disappearance compensates for the simultaneous appearance of the
other pole. Thus, with one of the accretion regions always visible,
and with spin-phase varying absorption playing little role in this star,
the spin pulse has a low amplitude and a complex shape, probably dependent
on slight asymmetries between the poles. The eclipse study also shows
that the upper pole appears at phase 0.25 and disappears at phase 0.75, on
the phasing adopted in this paper and Hellier (1997a).

\subsection{Hiding the lower pole}
In outburst the pulse amplitude jumps to 90 per cent. Thus, either the outburst
alters the balance between the poles such that one becomes 10 times brighter
than the other, or one of the poles must be hidden. The former is unlikely,
since an accretion disc is expected to feed both poles of a dipole roughly
equally, regardless of whether or not it is in outburst. Any
asymmetry would be due to the structure of the field, and so would also
be present in quiescence. Hiding the lower pole, though, can occur owing to
the peculiarly high inclination of XY~Ari, as we now discuss.

If we assume that XY~Ari accretes in equilibrium during quiescence, and
thus that the disc is disrupted at the point where the Keplerian
frequency of the disc equals the rotational frequency of the white
dwarf (it cannot be much further out otherwise the centrifugal barrier
would inhibit accretion), we have, for a 206-s spin period and a \sqig1
\msun\ white dwarf, an inner disc extending inwards to 9 white dwarf
radii ($R$\lo{wd}). At the inclination of XY~Ari ($80 < i < 84$; Hellier 1997a)
the bottom of the white dwarf is only barely visible in quiescence (which,
indeed, placed the upper limit of 84\deg\ on the inclination).  

The increased mass accretion rate in outburst, however, will reduce the radius
of disc disruption according to $r \sim \dot{M}^{-2/7}$ (e.g.~Frank, King \&\
Raine 1992). Since there are no substantial spectral changes in outburst, other
than the increased absorption, we can take the unabsorbed bremsstrahlung flux
as an indicator of the relative accretion rates. Depending on the absorption
model the unabsorbed flux, and therefore $\dot{M}$, increases by a factor
\sqig 24 between quiescence and the peak of outburst (Fig.~6). This 
corresponds to a 60 per cent reduction in the inner disc radius, $R$\lo{mag}, 
to \sqig 4 $R$\lo{wd}\ (see Fig.~8).

\begin{figure*}\vspace{3.5cm}   % Fig 8
\includegraphics{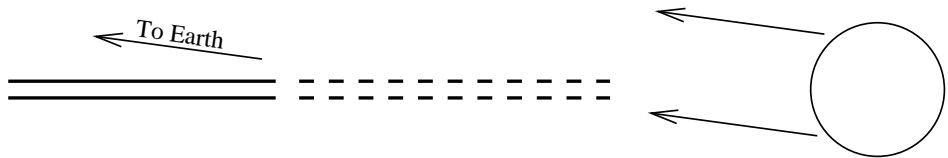}
\caption{A scale drawing of XY~Ari. The inner disc extends inwards to \sqig9
R\lo{wd}\ in quiescence and to \sqig 4 $R$\lo{wd}\ in outburst. The lower
accretion pole is visible in quiescence but not during outburst. The arrows are
drawn for an inclination of 82\deg.}
\end{figure*}

At $R$\lo{mag}\ = 4 $R$\lo{wd}\ the disc will obscure the white dwarf down to a
latitude of --19\deg\ (for $i$\,=\,84\deg) or --34\deg\ (for $i$\,=\,80\deg).
Hence, since the  lower accretion region has a latitude between --43\deg\ and
--63\deg\  (in quiescence) it will not be seen during the outburst. Although
the accretion regions will probably change latitude in outburst, this
calculation assumes zero disc thickness at the disruption point, and so in
reality more of the white dwarf will be hidden.

As a result the spin pulse will be close to total, with pulse maximum occuring
when the upper pole is on the visible face and a deep minimum occuring 
when it lies behind the white dwarf. This is indeed the phasing of the observed
pulse, which has flux maximum at phase 0.5, quarter of a cycle after the upper
pole appears at phase 0.25. Further the absence of spectral change over the
pulse is exactly as expected for a pulsation caused essentially by occultation.

\subsection{The spin pulse in outburst}
The residual flux at spin minimum cannot result from lower-pole emission
leaking through the disc, since this would be much more absorbed than the pulse
maximum spectrum, and no change in absorption is seen. Thus the
residual emission implies that the upper accretion region is sufficiently
extended that some of it is always on the visible face, even when the bulk 
has passed over the limb.

Indeed, there are two further indications of this. First, if the upper
accretion region lay, at some phases, entirely on the visible face or
entirely on the hidden face, it would produce a flat-topped or flat-bottomed
pulse. However the observed pulse near the peak of the outburst is sinusoidal.
Hence, parts of the upper accretion region must be continually appearing and
disappearing at all spin phases.

The emission site is expected to be an arc extending along a ring of
approximately constant magnetic latitude around the magnetic pole, with a
gradient in accretion rate along it so that maximum accretion occurs furthest
from the spin axis. This is discussed further in Hellier (1997a), who showed
that the offset angle of the magnetic axis from the spin axis is $8^{\circ} <
\delta < 27^{\circ}$ and that the magnetic colatitude of the accretion
regions, $\epsilon$, is \sqig19\deg\ in quiescence. In outburst, the decrease
in $R$\lo{mag}\ increases the magnetic colatitude of the accretion, so
that we expect $\epsilon \sim 30$\deg\ for accretion from 4 $R$\lo{wd}. Part of
the ring will always be in the visible hemisphere if $\epsilon > \delta -
(90\!-\!i)$, which is almost certainly satisfied in outburst.

In order to explain the sinusoidal pulse at the peak of outburst we then
require an accretion arc extended sufficiently along the ring to ensure that
some of the accretion is always in the visible hemisphere. There is no simple
analytical expression for this requirement, but the arc length must exceed 
180\deg\ unless $\delta < (90\!-\!i)$. This means that the pole picks up at
least some material from the disc at azimuths at which the majority of the
material is flowing to the other pole. This is energetically possible if the
disc is thickened out of the disc plane. 

Notice, though, that during the outburst rise and also during the decline 
the pulse profile is flat-topped. Thus we suppose that the spread of accretion
round the ring is not as extensive during the rise and the decline, and thus
that the range of azimuths over which accreting material is picked up is a
function of the accretion rate. 

The second indication that part of the upper pole is always visible comes from
the eclipse egress timings presented in Hellier (1997a). The egress times will
depend on the location of the accretion region on the white dwarf, and thus on
spin phase. The quiescent egresses can be divided into those of the upper pole
(which occur earlier, since the angle of the secondary star limb causes the
upper part of the white dwarf to emerge slightly earlier) and those of the
lower pole (which occur later). At the peak of the outburst (observation 6) the
egress occurs when the upper pole should be on the hidden face of the white
dwarf, but it occurs too early to result from the emergence of the lower pole. 
Thus it again implies that, at least in outburst, part of the upper pole 
is always visible, even when the bulk is on the hidden face.

In the quiescent observations (Hellier 1997a) we used the eclipse egresses to
estimate the size of the accretion regions, finding that they are  $<$0.002 of
the white dwarf surface. We cannot easily make a similar estimate in outburst
because the three observations at the peak of outburst all emerge from eclipse
near spin minimum, when the count rate is too low to make a reliable estimate
of egress duration, and further,  because one of  the three suffers a data
drop-out just at egress. The preceding discussion implies that the accretion
arcs are greatly extended in magnetic longitude compared to quiescence, but
since we have no estimate of their width (extent in magnetic latitude) we
can't estimate their fractional area, $f$. A complete ring with $\epsilon \sim
30^{\circ}$ would enclose an area of $f$ = 0.067, and this number is important
in determining the shape of the lightcurve, but the accretion will only occur
around the perimeter of this area.

Having explained the spin pulse in outburst we can further explain the 
slight rise in hardness ratio at pulse maximum. Some of the X-rays from the 
shock will be reflected by the white dwarf and these will have a harder
spectrum, since the softer photons are preferentially absorbed in the 
white dwarf atmosphere (e.g.~Done, Osborne \&\ Beardmore 1995). The reflected
component will be seen preferentially when the accretion area is face on, that
is at spin maximum, and so produce the increase in hardness ratio. The same
effect is seen in AM~Her stars when the accretion region faces us
(Beardmore \etal\ 1995).

In addition to explaining the enhanced spin pulse during outburst the above
picture also accounts for the increased absorption. There is substantial 
observational evidence that discs in DNe thicken during 
outburst, producing absorbing along lines-of-sight at angles up to \sqig20\deg\
above the plane (e.g.~Mason \etal\ 1988).  At the high inclination of XY~Ari 
this ensures that the flux from the upper accretion region will encounter the 
extra absorption seen in the spectra.

\subsection{The change in \mdot}
To check the plausibility of the interpretation we can make order of magnitude
estimates of the accretion flow, \mdot. At the peak of the outburst the
accretion rate is 24 times that in quiescence (scaling it by the normalisations
of the X-ray spectral fits, and assuming that only the upper pole is seen in
outburst). The increased \mdot\ through the inner disc means that the ram
pressure ($\propto$ \mdot) overwhelms the magnetic pressure at the original
magnetosphere. Thus only a fraction of the increased material accretes
immediately (the X-ray normalisations rise by  a factor 4; Fig~6) and most of
it pushes the disc inwards from \sqig 9 R\lo{wd}\ to \sqig 4 R\lo{wd}\ where
the magnetic pressure ($\propto r^{-6}$)  is sufficient to  establish a new
equilibrium. Here, the flow from the inner disc again matches the flow into
it and the luminosity reaches 24 times the quiescent value. 

The delay of 0.3--2 d (Fig.~1) is the time taken to fill the disc between
\sqig9 and \sqig 4 R\lo{wd}. With a quiescent accretion rate of 3\pten{16}\
g\,s\minone\ (Warner 1997) it implies a mass of \sqig 5\pten{22}\ g for this
section of disc.  Using an outer disc radius of \sqig60 R\lo{wd}\ (Allan \etal\
1996), and assuming a constant surface density, the total mass of the disc is 
3\pten{24}\ g. Further, again scaling from the X-ray luminosities, the total
mass accreted over the 5 days of outburst is 1.5\pten{23}\ g.

For comparison, the theoretical disc by Ichikawa \&\ Osaki (1992), adjusted 
for the parameters of U~Gem, has a total mass of 7\pten{23}\ g and deposits
8\pten{22}\ g onto the white dwarf each outburst. A model by Cannizzo,
Wheeler \&\ Polidan (1986) has a total mass of 2\pten{24}\ g and again
deposits 8\pten{22}\ g per outburst. These values agree with our values 
to within the observational uncertainties. 

\subsection{The story of the outburst}
We summarise the above as follows. In quiescence the disc is disrupted
at \sqig9 $R$\lo{wd}, allowing a clear view of the white dwarf. Accretion
occurs roughly equally onto both poles. The appearance/disappearance
of one pole compensates for the disappearance/appearance of the opposite
pole, producing a low-amplitude pulse as the white dwarf spins.

Day 1: A disc instability occurs somewhere in the disc, feeding material
towards the magnetosphere at an increased rate. The magnetosphere shrinks and
the inner disc occults the lower pole. Left alone, the upper pole now produces
a large amplitude spin pulse as it cycles  from the visible face to the hidden
face. Some of the increased mass flow reaches the white dwarf and the intrinsic
X-ray emission increases by a factor 4. However, since we are no longer seeing
the lower pole, and since the upper pole suffers increased absorption, the
observed count rate drops. The material accretes onto part of the ring around
the white dwarf, and at some spin phases the accreting region is entirely on
the visible face, producing a flat topped spin pulse.

Day 2: The mass flow onto the white dwarf increases dramatically, producing a
24-fold increase in X-ray emission over quiescence. Thus, even with the
increased absorption, the observed average count rate becomes three times that
in quiescence. Since the material is now connecting to field lines much further
in (\sqig 4 $R$\lo{wd}) the magnetic colatitude  of the accretion region climbs
from \sqig19\deg\ to \sqig30\deg.  Accretion also arrives from a greater range
of azimuth,  extending the accretion arc around the pole towards a complete
ring. Parts of the extended ring are therefore always disappearing and
appearing over the limb of the white dwarf and this, coupled with a gradient in
accretion rate along the ring, produces a sinusoidal spin pulse. 

Days 3 \& 4: The accretion rate and the observed count rate decline gradually.

Day 5: The accretion rate drops further and the magnetosphere expands.
The absorption decreases and so the observed count rate increases,
despite the lower accretion rate. The lower pole begins to appear again
and the pulse amplitude drops, with flux from the lower pole
filling in the minima. The accretion ring at the upper pole shrinks and
again lies entriely on the visible hemisphere at some spin phases; thus
the spin pulse again becomes flat topped. 

Day 6: XY~Ari returns to quiescence.

\subsection{Modelling the spin pulse}
To back up the results so far we present some simple modelling of the accretion
regions and thus of the spin pulse profiles. The model follows that
by Kim \&\ Beuermann (1995), tracing magnetic field lines back from the
disc disruption radius, $R$\lo{mag}, to their location on the white dwarf. 
The geometry is completely specified by $R$\lo{mag}, $\delta$ and $i$, and by 
assuming a dipolar field and a negligible shock height.

As the accretion rate is likely to be greatest along `downhill' fieldlines, we
have introduced a function, $A \pm B \cos\alpha$, where $\alpha$ is the angle
between the tangent to the field and the disc at the capture point. This
weights the intensity of the X-ray emission around the rings, using opposite
signs of $B$ for the upper and lower pole respectively. Depending on the
choice of $A$ and $B$ the emission can occur from all points on the ring
($A=0$), a 180\deg\ arc ($B=0$), or somewhere inbetween. The code sums the
flux from the accretion regions not obscured by the white dwarf, ignoring
absorption effects.

Fig.~9 shows simulations with the parameters adjusted to match the XY~Ari
pulse profiles (Fig.~3). The bottom curve uses parameters suitable for 
quiescence ($i$ = 82\deg, $\delta$ = 15\deg, $R$\lo{mag}\ = 9\,R\lo{wd}) and
assumes that both poles are visible and are equally bright. If such a model is
entirely symmetric no net modulation is seen. The low-amplitude modulation
seen in quiescence requires symmetry breaking, for instance a non-zero shock
height of order 0.01 $R$\lo{wd}\ or a dipole placed off center by \sqig0.05
$R$\lo{wd}. Thus for the bottom curve the dipole was offset by 0.05 
$R$\lo{wd}\ in a direction perpendicular to the plane formed by the spin axis
and the (centred) dipole axis. This reproduces the small peak at phase 0.7
(Fig.~3).

\begin{figure}\vspace{12cm}   % Fig 9
\includegraphics{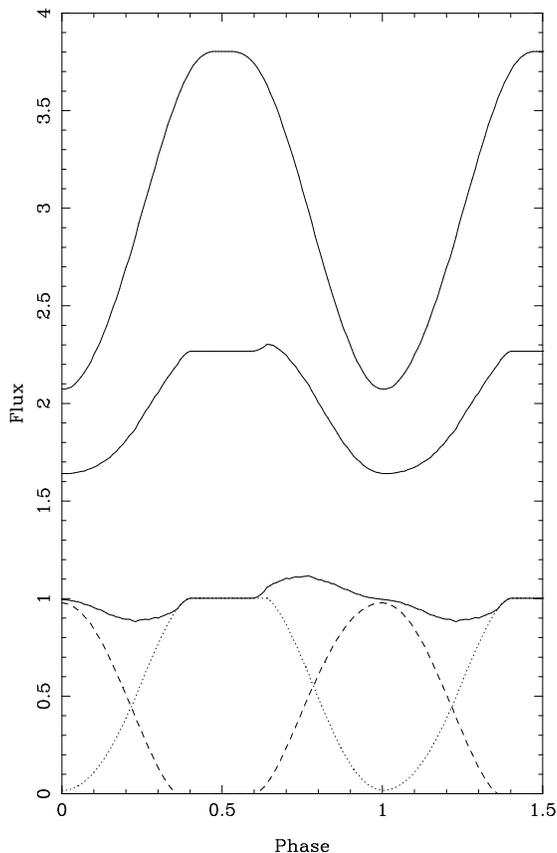}
\caption{Model spin pulse profiles with parameters adjusted for outburst, 
decline from outburst, and quiescence (top to bottom respectively). See
Fig.~3 for comparable data. The profiles are all normalized to 1, and the 
middle and upper profiles are offset by 1 and 2 units respectively.
For the quiescent profile the contributions of the upper pole (dotted line)
and lower pole (dashed line) are also shown.}
\end{figure}

The top profile in Fig.~9 uses the reduced $R$\lo{mag}\ of 4 $R$\lo{wd}\ and
assumes that only the upper pole is visible, as deduced for the peak of
outburst.  In order to obtain the \sqig90 per cent modulation depth the arc 
length has to be \sqiggt 210\deg.

The middle profile uses $R$\lo{mag}\ = 7 $R$\lo{wd}; this reduces $\epsilon$
compared to the outburst peak, and so the upper pole lies entirely on
the visible face of the white dwarf at some spin phases, producing
a flat-topped profile. We have assumed that the lower pole is partially
visible, and given it a 50 per cent weighting compared to the upper
pole. This fills in the mimima, producing a pulse profile comparable to
that during the decline from outburst (the slight asymmetry in the curve
arises because the off-centered dipole was used for all three curves).

Figs.~9 \&\ 3 show that the model reproduces the gross features of the 
data. There are some discrepancies, for instance the flat-topped section
of the pulse observed during the decline lasts longer than in the model.
However, several inputs to the model, such as the choice of dipole asymmetry
and the weighting of accretion rate with azimuth, are very uncertain.

\section{A possible alternative?}
Despite the above there is an alternative  interpretation which we can't rule
out. As outlined in Section~2,  at least some outbursts in IPs may be caused by
secondary star instabilities. Further, Hellier \etal\ (1989) suggested that 
during  an outburst of EX~Hya the increased mass transfer stream overflowed the
initial impact with the accretion disc and continued on a ballistic trajectory
to an impact with the magnetosphere. This could cause material locked in
orbital phase to feed onto field lines rotating with the spin cycle, and so
produce an accretion geometry varying with the beat (spin--orbit) frequency.
Thus, Hellier \etal\ (1989) predicted that EX~Hya in outburst should show an
X-ray beat pulse. Although no such observation has yet been made 
this idea seems to apply to other IPs in quiescence (Hellier 1993b; 1997b).

Unfortunately, since the current data on XY~Ari are sampled over
such a limited range of orbital phase (15 min near egress from a 
6 h orbit), there is no way of telling whether the pulsation is at the
spin period or the beat period. Thus the large changes in the pulsation
from quiescence could be explained if a stream overflowing in outburst
causes a beat period pulsation which is not present in quiesence.

We consider this less likely, though, for three reasons. First, XY~Ari has 
both a wide orbit and a short spin period compared to other IPs, and therefore 
a small magnetosphere. A ballistic trajectory (see e.g.~Lubow 1989) would have
a closest approach to the white dwarf of  \sqig9\pten{9}\ cm, outside the
magnetosphere of \sqig5\pten{9}\ cm. Thus it would probably interact with the
disc rather than the magnetic field (see Hellier 1997b for more discussion of
this point amongst IPs). Secondly, the other aspects of the outburst, which fit
into the disc-instability picture outlined above, do not find a ready
explanation in this picture. Thirdly, the probable `mass-transfer' outbursts
in EX~Hya, TV~Col and V1223~Sgr have several differences from normal outbursts,
including being very short at 1--2 d duration. The present outburst of XY~Ari
is longer at 5 d, which is more typical of normal DN events. Thus we do
not adopt the disc-overflow idea in interpreting XY~Ari, but an X-ray
lightcurve round a full orbital cycle in outburst would distinguish the
spin period from the beat period, and so settle the issue. 

Even if the changes in the pulsation are caused by the inner disc pushing
inwards, rather than by disc overflow, it is conceivable that the extra
material could still originate from a mass transfer event, rather than a disc 
instability, but we can't distinguish between these possibilities with our
current data.

\section{Discussion}
Our {\it RXTE\/} observations of XY~Ari over a 30-d period show an outburst
lasting 5 d. The intrinsic X-ray luminosity at the peak of outburst was 24
times that in quiescence. On the dubious assumption that the optical flux
increased by the same amount the optical amplitude would have been \sqig 3 mag,
typical of DNe. The recurrence time is much harder to estimate. This is
the first outburst to been seen, while, in addition to our {\it RXTE\/} data, 
XY~Ari has been observed with {\it Einstein\/}, {\it Ginga\/} and {\it ASCA\/}
on 7 occasions and in the infra-red on at least 5 occasions, counting an
`occasion' as a period of \sqig 5 d (for references see the introduction). Thus
we can only conclude that the time spent in quiescence is \sqiggt 10 times the
length of outburst.

During outburst the large, sinusoidal X-ray spin pulse is caused by an 
upper accretion pole disappearing and reappearing over the white 
dwarf limb as the white dwarf rotates. This behaviour is similar to the
model of IP spin pulses proposed by King \&\ Shaviv (1984).
Although we envisage the accretion region to be a ring surrounding the
upper magnetic pole, rather than the filled circle considered by King \&\ 
Shaviv, the effect will be similar as far as the pulse profile 
is concerned. Hellier, Cropper \&\ Mason (1991) criticized King \&\ Shaviv's
model on the grounds that it required that the lower pole be hidden,
otherwise symmetry would produce no net modulation. 
Hellier \etal\ showed that neither the disc nor the accretion flow could hide
the lower pole in a typical IP and concluded against the model as a general
explanation of the spin pulsations in IPs, instead looking to absorption to 
create the pulses.

Several circumstances combine in XY~Ari, though, to make the King \&\ Shaviv 
(1984) model apply during outburst. The inclination ($>$80\deg) is the highest 
amongst the known IPs; its unusually small spin period (206 s) 
implies a small magnetosphere; and the increased mass transfer in outburst 
further reduces the magnetosphere by half. The three effects combine to 
obscure the lower pole (see Fig.~8) in a manner which wouldn't occur in a
system with a lower inclination or a larger magnetosphere. As a further
indication of the different behaviour of XY~Ari in outburst, its X-ray spin
pulse is the only one to show increased hardness at flux maximum, whereas
virtually all other systems show a softer (less absorbed) spectrum at maximum
(e.g.~Ishida 1991). See Hellier (1997a) for further discussion of the 
spin pulse in XY~Ari and how it differs from that in other IPs.

GK~Per is the only other IP to have been observed in outburst in X-rays, 
and the similarity with XY~Ari is striking. GK~Per again has a short spin
period (351 s) and thus a small magnetosphere. It also has a low-amplitude
complex spin pulse in quiescence (e.g.\ Ishida \etal\ 1992),  which could
well imply an asymmetry between the poles. In outburst the pulse amplitude
is much larger and the pulse profile simpler  (Watson \etal\ 1985).  Clearly
the same mechanism of blocking the lower pole in outburst might apply to
GK~Per, which would require that it has at least a moderately high 
inclination.

\subsection{Application to `non-magnetic' systems}
Following from the above we can try to apply the idea to non-magnetic CVs
to explain dwarf novae oscillations (DNOs). These are quasi-periodic
oscillations seen both in the optical and X-ray in many DN during
outburst. Warner (1995b) reviews the observations and argues that they imply
fields of 10$^{4}$--10$^{5}$ G, below those of IPs but sufficient to
channel the accretion flow close to the white dwarf. The lower coherence of the
oscillations results from the surface layers of the white dwarf slipping under
the influence of accretion, whereas the higher field in IPs binds the surface
to the core.

While explaining much of the DNO phenomenology this doesn't explain why the
oscillations are seen preferentially in outburst. In fact, since the 
magnetospheres will be larger in quiescence, one might expect better
channelling and larger pulsations in quiescence. If again, though, both poles
are seen in quiescence, but the disc pushes inwards to obscure the lower pole
in outburst, the broken symmetry will lead to larger pulsations in outburst.

Systems with B \sqig 10$^{4}$--10$^{5}$ G will have smaller magnetospheres
than XY~Ari, but, in general, the inclinations will also be smaller, allowing
the same transition between viewing two poles and viewing one pole when the
accretion rate changes. 

Such fields have additionally been proposed as an explanation for the UV delay
seen in some DNe, where the UV flux rises \sqig12 h after the optical
flux (e.g.~Hassall \etal\ 1983). Such a field would empty the inner (UV
emitting) part of the disc in quiescence. Livio \&\ Pringle (1992) propose that
at the start of an outburst the disc material has to refill the depleted region
before accreting, explaining the delay in the rise of the higher energy flux.
The fact that we see exactly that in XY~Ari supports this idea. In our data the
increased absorption and blocking of the lower pole, indicating changes in
the disc, are seen the day before the major rise in X-ray flux.
During this time the inner disc is diffusing inwards and collapsing the
magnetosphere (Section 5.3).

\section*{Acknowledgments}
We are grateful to the {\it RXTE\/} team for their execution of a difficult
observation and for the assistance provided by the e-mail help desks.

\end{document}